\documentclass[floatfix,aip,amsmath,amssymb,reprint]{revtex4-1}

\usepackage{epsfig}
\usepackage{latexsym}
\usepackage{amsmath}
\usepackage{amsfonts}
\usepackage{amssymb}
\usepackage{amsbsy}
\usepackage{subfigure}
\usepackage{bm}
\usepackage{color}
\usepackage{dcolumn}
\usepackage{hyperref}

\usepackage{graphicx}

\newcommand{\lla}{\left\langle}
\newcommand{\rra}{\right\rangle}

\begin{document}

\title{Phase segregation of
  liquid-vapor systems with a  gravitational field} 

\author{A. Lamura}
\email[]{antonio.lamura@cnr.it}
\affiliation{Istituto Applicazioni Calcolo, CNR, Via Amendola 122/D,
  70126 Bari, Italy}

\date{\today}

\begin{abstract}
  Phase separation in the presence of external forces has attracted
  considerable attention since the initial works for solid mixtures.
  Despite this, only very few studies are available which address 
  the segregation process of liquid-vapor systems under gravity.
  We present here an extensive study which takes into account both
  hydrodynamic and gravitational effects on the coarsening dynamics.
  An isothermal formulation of a lattice Boltzmann model for a liquid-vapor
  system with the van der Waals equation of state is adopted.
  In the absence of gravity, the growth of domains follows a power law
  with the exponent $2/3$ of the inertial regime.
  The external force deeply affects the observed morphology 
  accelerating the coarsening of domains and favoring the liquid accumulation
  at the bottom of the system. Along the force direction, 
  the growth exponent is found to increase with the gravity strength
  still preserving sharp interfaces since the Porod's law is found to be
  verified. 
  The time evolution of the average thickness $L$
  of the layers of accumulated material at confining walls
  shows a transition  from an initial regime where $L \simeq t^{2/3}$
  ($t$: time) to a late-time regime $L \simeq g t^{5/3}$ with $g$ the
  gravitational acceleration.
  The final steady state, made of two overlapped layers of liquid and
  vapor, shows a density profile in agreement with theoretical predictions.
\end{abstract}


\maketitle 

\section{Introduction}
\label{sec:intro}

The quench of an initially mixed liquid-vapor system
to a state with two coexisting phases
enables the formation of domains with different composition.
This is a well known
problem which has attracted a lot of attention in
the past years
\cite{yamamoto1994,osborn1995,Kabrede2006,swift1996,sofonea2004,
    lamorgese2009,ganPRE2011,das2011,Roy2012,Midya2017,sofonea2018,
    zhang2019,Negro2024,
    Para2025}.
The main interests lie
in both the comprehension
of the mechanisms behind the formation and growth of domains
and the application in the industrial
processing of soft materials.

Theoretical studies, numerical simulations, and experimental
  investigations addressed the phenomenon of phase separation
  \cite{bray1994,onuki2002,cates2017}. There is convincing evidence
  that growth of domains occurs in a self-similar way.
  Domain patterns at later times
    look, in a statistical sense, similar to those at earlier times,
    apart from a change of scale. This leads to the introduction of the
    scaling hypothesis.
The idea can be formalized by introducing a single
characteristic domain size $R$
which grows in time with a power law $R \sim t^{\alpha}$ ($t$: time).
The growth exponent $\alpha$ assumes different values depending on the
physical process governing the phase separation
as outlined in the following. 
We will restrict our discussion to the case of liquid-vapor
systems
with symmetric composition
  (equal fractions of liquid and vapor).

At early times, when hydrodynamics is not effective,
the growth dynamics shows an exponent $\alpha=1/2$,
  as predicted for nonideal
  fluids \cite{Koch1983}.
  This exponent is due to a different realization
  of the Lifshitz-Slyozov mechanism where the transport of molecules
is kinematic rather than diffusive \cite{onuki2002}.
This result is of more general validity holding
for liquid-vapor-type phase separation
in simple and complex fluids far from the critical point \cite{tateno2021}.
Hydrodynamics is operative at late
times and modifies the growth exponent. An intermediate viscous
hydrodynamic regime with exponent $\alpha=1$ is then followed at long times
by the inertial hydrodynamic regime with $\alpha=2/3$, as it can be
derived also by dimensional analysis \cite{bray1994}.
The viscous regime is due in three dimensions by the Rayleigh-Plateau
instability that induces the breaking of connected domains so that
retracting broken channels favor pattern swelling \cite{bray1994}.
This mechanism is inoperative in two dimensions \cite{yeomans2000}.
Finally, the inertial regime is driven by the Laplace pressure
mismatch between the outer and inner parts
of a domain which promotes more rounded bubbles \cite{bray1994}.

The exponent $\alpha=1/2$
has been found in molecular dynamics (MD)
simulations \cite{yamamoto1994} as well as in lattice Boltzmann (LB) models
both in two
\cite{osborn1995,swift1996,sofonea2004,CRISTEA2006113,Gonnella2007,ganPRE2011,zhang2019}
and three dimensions \cite{Negro2024}.
The viscous regime is present only for binary fluid mixtures and
  not for liquid-vapor systems. Indeed,
a recent detailed LB study \cite{Negro2024} has ruled out
its existence in three-dimensional
isothermal phase separation of liquid-vapor systems by considering a
universal curve \cite{kendon2001} for the different scaling regimes:
the early exponent $\alpha=1/2$ is followed by the inertial
exponent $2/3$ at long times.
Lattice Boltzmann simulations
\cite{osborn1995,swift1996,sofonea2004,ganPRE2011,zhang2019,Negro2024} 
and the direct numerical solution of the Navier-Stokes equation for a van der
Waals (vdW) fluid \cite{lamorgese2009} find evidence only of the exponent
$\alpha=2/3$ at late times in two and three dimensions. 
For completeness we report that in three-dimensional
  MD simulations of liquid-vapor systems,
  it was found a linear growth exponent at long times
  \cite{das2011,Davis2025}.

Though a well detailed quantitative description of the coarsening
phenomenology is available, not so many studies have been conducted to address
the impact of an external field when a liquid-vapor system undergoes
phase separation.
The role played by an imposed shear flow 
has been considered by using LB simulations
in Refs.~\onlinecite{wang2009,lin2025} where it is outlined
that the competition between surface tension and shear force produces
either
a single liquid droplet when the surface tension dominates or
an horizontal stripe (band) of liquid spanning the system along the
flow direction when the shear force prevails.
The action of gravity deeply affects the morphology of a fully separated
system 
leading to stratification due to the density difference between the two
phases \cite{Fogliatto2019,Czelusniak2023}.

In order to
characterize quantitatively
the phase separation with gravity,
 we perform LB simulations on very large systems ($2048^2$ nodes).
 We implement a two-dimensional LB model \cite{coclite2014}
which is obtained by using a Gauss-Hermite projection
of the Boltzmann equation \cite{shan2006bis}.
The resulting LB equation is supplemented by a body force \cite{Guo2002}.
This latter one includes both the external contribution from a force and
the internal one
derived from the vdW free-energy functional \cite{widom},
which ensures that the vdW equation of state is locally satisfied
by the fluid.
It is possible to prove that the continuity and the generalized Navier-Stokes
equations without spurious terms can be recovered in the continuum limit.
Moreover, the surface tension is included in the pressure
tensor and can be set without affecting the equilibrium phase diagram
of the system.
This model has been used to correctly describe the behavior of a
liquid-vapor system subject to a periodic potential under the action
of a shear flow \cite{coclite2014} and to provide reliable quantitative
results in the problem of cavitation inception at a
sack-wall obstacle \cite{kaehler2015}.

We perform a systematic study of the effects due to gravity
in the phase separation process in the
inertial regime.
This is motivated by the fact that a model like ours, which includes
  hydrodynamic interactions,
  is more realistic from an experimental point of view
\cite{Davis2025}.
We find
evidence that the external force deeply affects the growth rate of forming
domains. Their average size along the
force field grows
with a power-law behavior whose exponent is larger than the value
$2/3$ observed in the absence of gravity. The growth exponent
increases with the external force
up to the value $\simeq 5/3$ for the largest gravity strength
used in our study.
During the segregation process, liquid sediments at the bottom
so that in the steady state two overlapped
layers of liquid and vapor span completely
the system.
The time evolution of
the average thickness $L(t)$ of layers of material accumulated at
bottom and top of the system, shows
a transition
from an initial regime where $L(t) \sim t^{2/3}$
to a late-time behavior
characterized by the law $L(t) \sim g t^{5/3}$, $g$ being the
gravitational acceleration. The Porod's law \cite{oono1988}
is shown to be verified along the
force field and the numerical
steady density profile is found in excellent agreement
with the theoretical one \cite{Santos2002}. 

Very few studies are available
which consider the
coarsening dynamics of liquid-vapor systems
  under gravity \cite{articlecristeagravity,Davis2025}.
Anisotropy in forming domains has been observed in a LB study
in two dimensions with a growth exponent $\alpha=1$ in the
gravitational direction, not depending on the force magnitude
\cite{articlecristeagravity}. The smaller size of the
lattice used in such simulations
does not allow the authors to follow the sedimentation process on
long times. This prevents an accurate evaluation of the time behavior
of the average thickness $L$ of sedimented layers.
Very recently, atomistic simulations 
have provided evidence of an accelerated growth along the gravitational
direction strongly depending on the force strength \cite{Davis2025}.
In this latter case, the authors find a transition to a growth exponent
$\alpha > 1$ when gravity acts on the phase separation process. This
transition occurs at earlier times when increasing gravity and it is
found that $\alpha \rightarrow 3$ in the limit of the highest considered
value of the external force. The difference in the observed
exponents with respect to the present results might be attributed to the adopted
numerical method.
Our study is the first one that can
  investigate the influence of gravity on the unexplored
  inertial regime.

The paper is organized as follows. Section II presents
the LB model used
  in the present study with the simulated continuum equations.
  Then, we discuss the phase separation process in the presence of gravity
  at low viscosity and characterize quantitatively the separation process.
  Finally, we conclude with a discussion of the main findings of the present
study.

\section{Numerical model}
\label{sec:model}

We briefly sketch here the two-dimensional isothermal
LB model used in this work.
Details on the numerical method and its implementation can be
found in Refs.~\onlinecite{coclite2014,kaehler2015}.

A set of nine distribution functions $\{f_i({\mathbf r},t)\}, i=0,\dots,8$, is
introduced on the nodes of a lattice.
This set is associated to the lattice velocities
$\{{\mathbf e}_i\}, i=0,\dots,8$, at the lattice site ${\mathbf r}$
and discrete time $t$ with time step $\Delta t$.
Here we consider the so-called $D2Q9$ square
lattice where each site is connected to its nearest and next-to-nearest
neighbors located at distances $\Delta s$ and $\sqrt{2} \Delta s$, respectively.
The distribution functions evolve in time according to the LB equation
\cite{HuangSukopLuMultiphaseLB,krueger17,succi18}
\begin{eqnarray}
  f_i({\mathbf r}+{\mathbf e}_i \Delta t, t+\Delta t)-
  f_i({\mathbf r}, t) = && -\frac{\Delta t}{\tau}\big[f_i({\mathbf r}, t)
    -f_i^{eq}({\mathbf r}, t)\big] \nonumber \\
  &&           + \Delta t F_i , \;\; i=0,\dots,8 .
\end{eqnarray}
The forcing terms $\{F_i\}$ have to be properly determined,
$\tau$ is the relaxation time, and
$\{f_i^{eq}({\mathbf r},t)\}$ are the equilibrium distribution
functions. They are fixed by a second-order Hermite expansion
of the discrete Maxwell-Boltzmann distribution function
\cite{shan2006bis,coclite2014}
so that it results
\begin{eqnarray}
  f_i^{eq}(n, {\mathbf v}, T)&=& \omega_i n \Big\{
  1+e_{i,\alpha} v_{\alpha}
    +\frac{1}{2}\big[v_{\alpha}v_{\beta}(e_{i,\alpha}e_{i,\beta}-\delta_{\alpha \beta}) \nonumber \\
  &+& (T-T_c)(e_i^2-2)\big] \Big\} ,
\end{eqnarray}
where $n=\sum_i f_i$ is the fluid density,
$n {\mathbf v}=\sum_i f_i {\mathbf e}_i + \Delta t {\cal F} / 2$ is the
fluid momentum, ${\cal F}$ being the force density related to
$\{F_i\}$, $T$ is the temperature
which is kept fixed in our model,  and $T_c$ is the critical temperature of the
liquid-vapor system.
The Greek index denotes the spatial direction.
The Hermite expansion sets the moduli $|{\mathbf e}_i|=\Delta s / \Delta t
= \sqrt{3}$ for the horizontal and vertical links ($i=1,\dots,4$),
$|{\mathbf e}_i|= \sqrt{6}$ for diagonal links ($i=5,\dots,8$),
$|{\mathbf e}_0|= 0$ for the rest velocity, and the weights $\omega_i=1/9$
($i=1,\dots,4$), $\omega_i=1/36$ ($i=5,\dots,8$), $\omega_0=4/9$.  

The forcing term $F_i$ is expressed as \cite{Guo2002,coclite2014}
\begin{eqnarray}
  F_i&=& \omega_i (1-0.5 \Delta t/\tau)\Big\{e_{i,\alpha}{\cal F}_{\alpha}
  \nonumber \\
  && +0.5 \big[v_{\alpha}{\cal F}_{\beta}+v_{\beta}{\cal F}_{\alpha}
  +(T_c-T) \nonumber \\
  && \times (v_{\alpha}\partial_{\beta}n+v_{\beta}\partial_{\alpha}n+
  \partial_{\gamma}(n v_{\gamma})\delta_{\alpha \beta})\big] \nonumber \\
  && \times (e_{i,\alpha}e_{i,\beta}-\delta_{\alpha \beta})\Big\} ,
\end{eqnarray}
where ${\cal F}$ is the sum of external contributions ${\cal F}_e$ and
of internal ones ${\cal F}_i$.
In the following we will use
\begin{equation}
{\cal F}_e=(0,-ng)  
\end{equation}
representing a force along the $y$-direction directed downwards with $g$
the gravitational acceleration. Willing to describe a vdW fluid, it can
be shown that the internal force has to be \cite{coclite2014}
\begin{eqnarray}
  {\cal F}_{i,\alpha}&=&\partial_{\alpha}(p^{id}-p^{vdW})+\kappa n \partial_{\alpha}
  \nabla^2 n \nonumber \\
  &=& \partial_{\alpha} p^{id}-\partial_{\beta} \Pi_{\alpha \beta} ,
\end{eqnarray}
where $p^{id}=nRT$ is the ideal equation of state (EOS),
$p^{vdW}=3 n R T/(3-n)-9 n^2/8$ is the van der Waals EOS with critical point
located at $(n_c=1, T_c=1, p_c^{vdW}=3 R / 8)$, $R$ being the gas constant.
\begin{eqnarray}
  \Pi_{\alpha \beta}&=& \big[p^{vdW}-\kappa n \nabla^2 n -
    0.5 \kappa (\nabla n)^2\big]
  \delta_{\alpha \beta}\nonumber \\
  &+&\kappa  \partial_{\alpha} n \partial_{\beta} n 
\end{eqnarray}
is the pressure tensor \cite{widom} and 
the variable $\kappa$ controls the energy cost for interface formation
allowing the variation of the surface tension without modifying
the temperature.

It can be proved \cite{coclite2014}
that in the continuum limit the continuity equation
\begin{equation}
\partial_t n + \partial_{\alpha}(n v_{\alpha})=0
\label{eq:cont}
\end{equation}
and the Navier-Stokes equation
\begin{eqnarray}
  &&\partial_t(n v_{\alpha})+ \partial_{\beta}(n v_{\alpha} v_{\beta})=
  \nonumber \\
  &&-\partial_{\beta}\Pi_{\alpha \beta}+{\cal F}_{e,\alpha}
  +\partial_{\beta}\big[\eta(\partial_{\alpha}v_{\beta}
    +\partial_{\beta}v_{\alpha})\big]
\label{eq:ns}
\end{eqnarray}
are correctly recovered where $\eta=n(\tau-\Delta t/2)$ is the fluid viscosity.
The continuum approximation of hydrodynamics, as described
by the two previous equations, is
valid in the limit of negligible Knudsen number $Kn$.
This is the case in our model for which
$Kn= \Delta s \tau / (\Delta t W) < 10^{-3}$
with the choice made for the model
parameters (see the following), $W$ being the size of
the system.

In order to reduce spurious velocities at interfaces \cite{shan2006},
a 9-point stencil for second-order finite difference scheme is implemented
to compute spatial derivatives.
The values of the two free parameters in the stencil \cite{pooley2008_bis}
have been fixed to minimize the spurious velocities (see
Ref.~\onlinecite{coclite2014} for details).

In the following all the physical quantities are made dimensionless
by dividing them for reference quantities. Therefore we introduce the reduced
density $n_r=n/n_c$, the reduced temperature $T_r=T/T_c$, and
the reduced pressure
\begin{equation}
p_r^{vdW}=p^{vdW}/p_c^{vdW}=\frac{8 n_r T_r}{3-n_r}-3 n_r^2 .
\label{eq:p}
\end{equation}
The strength of the gravity field is expressed by the
dimensionless energy
\begin{equation}
  E_r=\frac{MgH}{RT_c}
  \label{eq:grav}
\end{equation}
where $M$ is the molar mass and $H$ is the system height.
In the following we will set $M=R=1$,
as usually done in the literature \cite{Fogliatto2019,Czelusniak2023}.

\section{Results}
\label{sec:results}

In all the simulations we consider a square lattice
with $N=2048$ nodes per direction with $\Delta s=1$ and
$\Delta t=\sqrt{3}/3$ to be consistent with the value of the lattice velocity
fixed by the Gauss-Hermite quadrature on the $D2Q9$ lattice.
Periodic boundary conditions (BC) are adopted
along the $x$-direction and
no-slip BC with local density conservation and neutral wetting
\cite{kaehler2015} are enforced at confining walls which are
placed along the lattice rows at $y=0, H$ with $H=(N-1) \Delta s$.
Moreover, we set $T_r=0.95<T_{r,c}=1$,
$\kappa=0.3$ in order to have the interface width \cite{wagner07}
$\chi = 2 \sqrt{2 \kappa T_r/(1-T_r)} \simeq 6 \Delta s$,
and $\tau=\Delta t$.
The considered value of the viscosity
is such to observe, in the absence of gravity,
full separation of the system within the inertial
regime recovering the scaling exponent $\alpha=2/3$ (see the following).
At the initial time, systems are
initialized in a symmetric mixed state with
density fluctuations of about $1\%$
around the mean value $n_r^{av}=(n_r^L+n_r^V)/2=1.021$
where $n_r^L=1.462$ and $n_r^V=0.579$ are
the coexisting densities of the
liquid and vapor phases, respectively,
at the chosen quenching temperature $T_r$,
as obtained from the Maxwell construction of the vdW EOS without gravity.
Gravity is applied
from the initial time in simulations
and
the value of the gravitational energy $E_r$ is varied in the
range $[0, 1.84] \times 10^{-2}$.
Results presented in the following sections
have been averaged over five realizations of the system.

\subsection{Phase separation dynamics}

At the beginning, the liquid and vapor phases start to separate by the formation
of domains with different composition
separated by sharp interfaces. In these early stages,
gravity does not introduce any visible anisotropic effect. Once
density is nearly constant in single-phase domains,
the evolution process appears to be influenced by the applied gravity
as well as by the presence of confining walls.

\begin{figure}[ht]
\begin{center}
\includegraphics*[width=\columnwidth,angle=0]{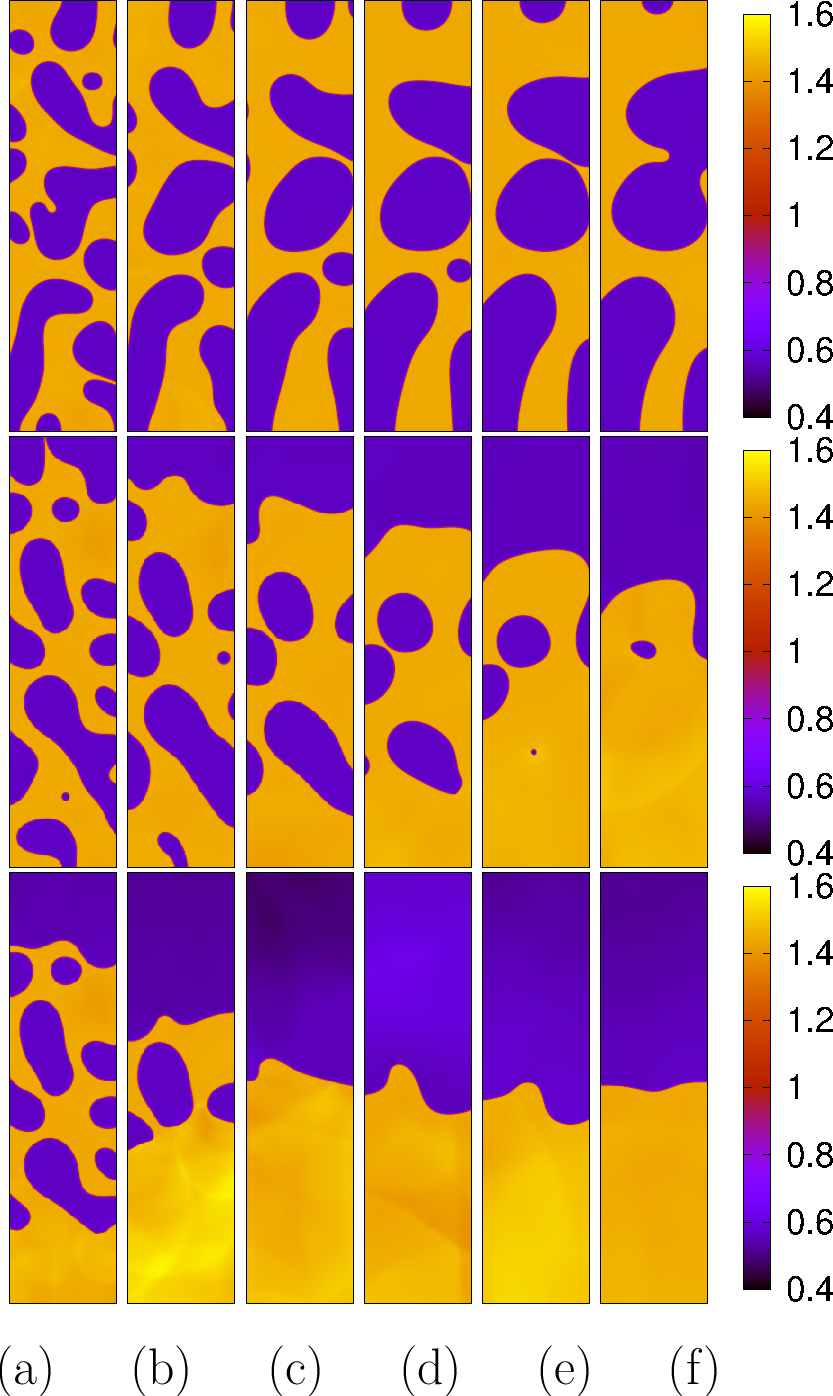}
\caption{Contour plots of the density $n_r$ in the cases with $E_r=0$ (top),
  $0.41 \times 10^{-2}$ (middle), $1.84 \times 10^{-2}$ (bottom)
  on a portion of size $(511 \Delta s \times 2047 \Delta s)$
  of the whole
  system at times
  $t/(5.2 \times 10^3 \Delta t)= 2$ (a), $3$ (b), $4$ (c), $5$ (d), $6$ (e),
  $7$ (f).
\label{fig:conf2d}
}
\end{center}
\end{figure}
\begin{figure}[ht]
\begin{center}
\includegraphics*[width=0.9\columnwidth,angle=0]{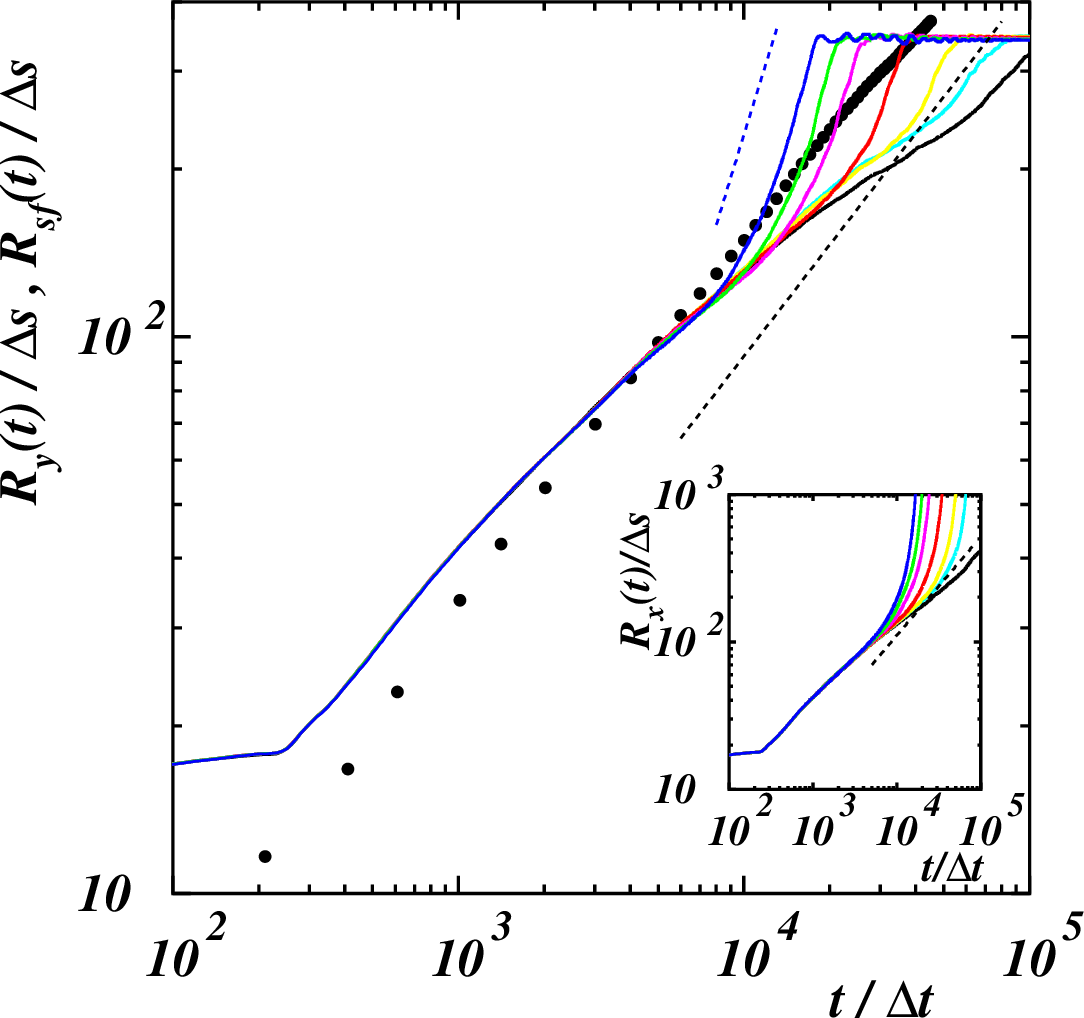}
\caption{Time evolution of the size of domains $R_y(t)$ along the vertical
  direction in the cases with
  $E_r= 0$ (black line),  $0.12 \times 10^{-2}$ (cyan
  line), $0.20 \times 10^{-2}$ (yellow
  line),
  $0.41 \times 10^{-2}$ (red line), $0.82 \times 10^{-2}$ (purple line),
  $1.23 \times 10^{-2}$ (green line),
  $1.84 \times 10^{-2}$ (blue line), and of the spherically-averaged domain
  size $R_{sf}(t)$ for the fully periodic system without gravity 
  (black dots).
  The black and blue dashed lines
  are guides to the eye and have slopes $0.66$ and $1.66$, respectively.
  Inset: Time evolution of the size of domains $R_x(t)$ along the horizontal
  direction in the same cases of the main panel. The black dashed line
  is a guide to the eye and has slope $0.66$.
\label{fig:raf}
}
\end{center}
\end{figure}

This can be seen in Fig.~\ref{fig:conf2d} where a sequence of
snapshots, taken on a rectangular portion of the whole system, is
shown in the absence of gravity (top row) and
for two values of the gravitational energy $E_r$.
The
Laplace pressure difference between the internal and external parts of domains
favors circular patterns, driving the fluid motion. The breakage of liquid
necks induces the breakup and reconnections of interfaces \cite{FURUKAWA1994}. 
This is evident in the bulk of system (see panels (c)-(f) of top and middle
rows of Fig.~\ref{fig:conf2d}). This is the leading mechanism driving the
phase separation and is responsible of the inertial growth exponent $2/3$
\cite{FURUKAWA1994}. 
However, the presence of confining walls modifies the domains morphology
in the surroundings. Indeed,
the constraint of neutral wetting induces normal interfaces thus inhibiting
the formation of circular bubbles (see patterns next to the bottom wall
in panels (c)-(f) of top row of Fig.~\ref{fig:conf2d}).
This process is responsible for the
slowing down of the growth rate without gravity in comparison to what can be
observed in a
system with periodic BC in both spatial directions
(see Fig.~\ref{fig:raf}).
On the other side,
domains are elongated in the vertical direction and reconnections
of vapor droplets are favored by the presence of the external force
(see panels (a)-(b) of middle and bottom rows of Fig.~\ref{fig:conf2d}).
Moreover, 
an effective sedimentation of the heavier phase at the bottom wall takes
place which
begins earlier as stronger values of gravity are used
(see panels (c)-(e) of middle and bottom rows of Fig.~\ref{fig:conf2d}).
The net effect is an increase of the growth rate with gravity
in the vertical direction (see Fig.~\ref{fig:raf}). 
The process continues until the two phases are fully separated
with a single smooth interface separating the liquid domain at bottom from
the vapor one at top of the system
(see panels (f) of middle and bottom rows of Fig.~\ref{fig:conf2d}).
We will describe in the following how the sedimentation occurs. 

In order to gain insight in the phase separation process, it is
useful to compute the structure factor
\begin{equation}
  C({\mathbf k},t)= \big(\tilde{n}_r({\mathbf k},t)\tilde{n}_r(-{\mathbf k},t)
  \big)
\end{equation}
where $\tilde{n}_r({\mathbf k},t)$ is the spatial Fourier transform of
$\big({n}_r({\mathbf r},t)-n_r^{av}\big)$.
This quantity allows us to determine the size of domains by computing
the inverse of the first moment of $C({\mathbf k},t)$. Due to the
inherent anisotropy of the present problem, we compute separately
the characteristic extensions
along the two spatial dimensions as
\begin{equation}
  R_{x,y}(t)=\pi\frac{\int d{\mathbf k} C({\mathbf k},t)}
  {\int d{\mathbf k} k_{x,y} C({\mathbf k},t)} .
\label{eq:raggi}
\end{equation}
The resulting measures are depicted in Fig.~\ref{fig:raf} for different
values of gravity.
In order to have a complete comparison with the isotropic system
without external force,
runs in this latter case have been also executed on fully periodic lattices.
The average size $R_{sf}(t)$
of domains is now computed
by using
the spherically
averaged structure factor
\begin{equation}
C_{sf}(k,t)= \lla  C({\mathbf k},t) \rra_k
\end{equation}
where the brackets $\lla \dots \rra_k$ indicate the average over
a shell at fixed $k$ in the reciprocal (discrete)
${\mathbf k}$-space. The values of $R_{sf}(t)$
are shown in  Fig.~\ref{fig:raf} 
and indicate a growth exponent fully consistent
with the value $\alpha = 2/3$ of
the inertial regime.
The analysis of the growth dynamics is performed at late times,
taking data after the formation of well-defined
  interfaces \cite{yeomans2000}.
  For the system size considered in the present study,
  approximately one decade of time is accessible.
As previously outlined, the growth in the absence of gravity
is slowed down by the presence of walls and slightly increases at long times
due to the observed reconnections of domains in the bulk
(see panels (e)-(f) of top row of Fig.~\ref{fig:conf2d}). This
behavior can be observed along both spatial directions.
The presence of interconnected patterns is as well verified in
fully periodic systems.
When gravity is enforced,
the initial growth of $R_y$ and $R_x$ is not affected by the external field
until time $t/\Delta t \simeq 8 \times 10^3$
for all considered values of gravity.
Then the fate of growth depends on the strength of the applied force.
At small values of $E_r$, the initial
growth along the vertical direction resembles
the one observed at $E_r=0$. Only at
later times gravity is effective
compensating the wall effect and
favoring
the downward motion of large domains with a crossover
to an accelerated growth in the $y$-direction
with exponent $\alpha > 2/3$.
This crossover is reached earlier
by increasing the applied external force
and the effective growth exponent increases up to
$\alpha \simeq 5/3$ for the largest value of the gravitational
energy $E_r$ used in our study. Once the system is completely divided in two
well separated domains
(see panels (c)-(f) of bottom row of Fig.~\ref{fig:conf2d}),
the sizes $R_y$ saturate to the same limiting values with damped oscillations.
The time to reach the final plateau decreases
with $E_r$ while the amplitude of oscillations is amplified by gravity.
The measured growth in the $x$-direction is extremely fast once the
sedimentation process sets in, and rapidly increasing with gravity.
This behavior is due to the periodic BC along this direction causing the
wrapping of the system on itself and the divergence of $R_x$
since the denominator of Eq.~(\ref{eq:raggi}) goes to zero.

\begin{figure}[ht]
\begin{center}
\includegraphics*[width=0.9\columnwidth,angle=0]{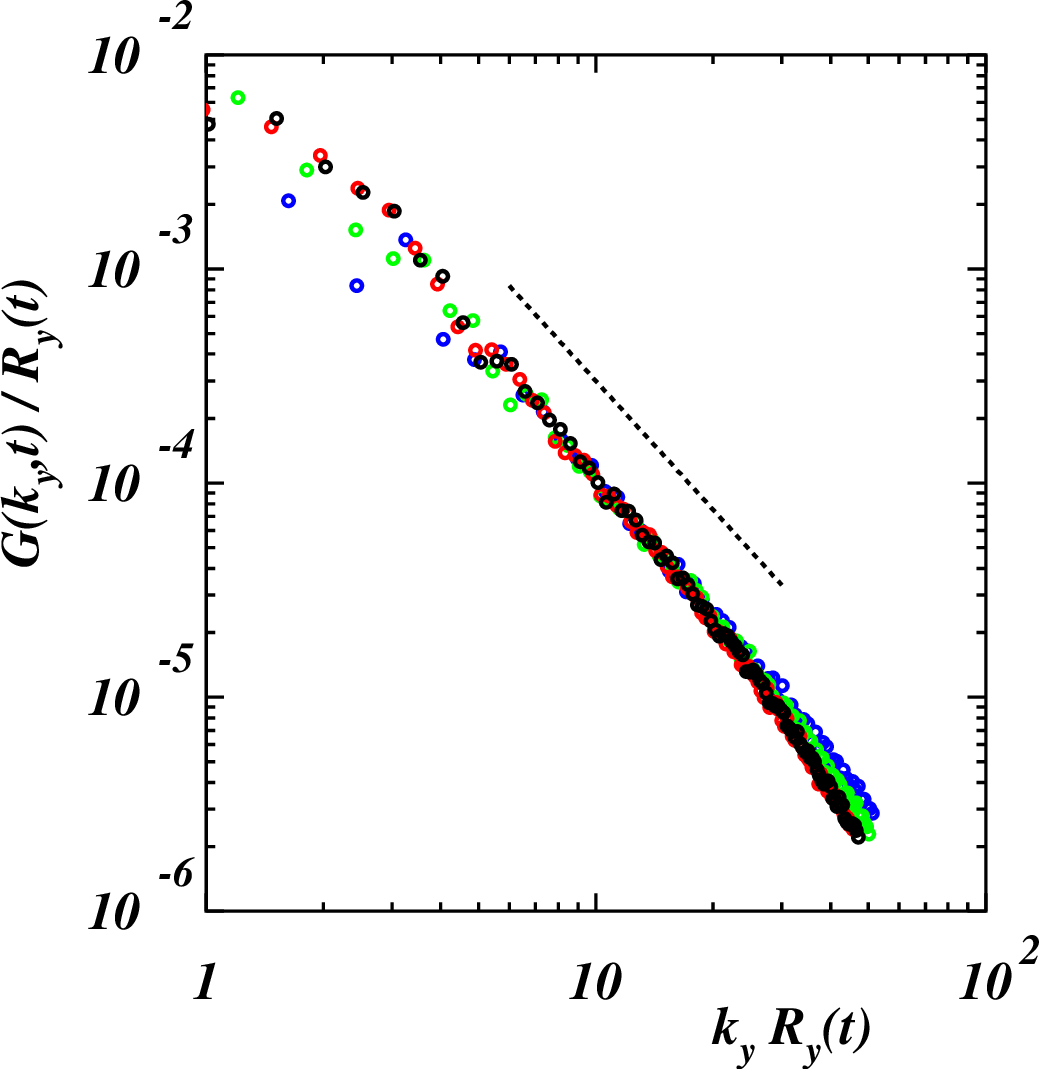}
\caption{Scaled structure factor $G(k_y,t)/R_y(t)$,
  averaged along the $k_x$-direction
  at time $t/(5.2 \times 10^3 \Delta t)= 3$, as a
  function of $k_y R_y(t)$ in the cases with $E_r= 0$ (black symbols),
  $0.41 \times
  10^{-2}$ (red symbols), $1.23 \times
  10^{-2}$ (green symbols), $1.84 \times
  10^{-2}$ (blue symbols). The dashed line has slope $-2$.
\label{fig:cdik}
}
\end{center}
\end{figure}

The structure of the system along the $y$-direction
can be investigated by considering the structure factor averaged
along $k_x$-direction in the reciprocal space
\begin{equation}
G(k_y,t)= \lla  C({\mathbf k},t) \rra_{k_x} 
\end{equation}
so to consider the system as effectively one-dimensional.
In the case of self-similar patterns in the gravity direction,
one would expect to verify the scaling behavior \cite{bray1994}
\begin{equation}
  G(k_y,t)= R_y \tilde{G}(k_y R_y(t))
 \label{eq:scaling} 
\end{equation}
where $\tilde{G}(k_y R_y(t))$ is  a time-independent scaling
function.
The scaled structure factors $G(k_y,t)/R_y(t)$ as functions
of the variable $k_y R_y(t)$ are shown in Fig.~\ref{fig:cdik}
for four different values of $E_r$ at time $t/(5.2 \times 10^3 \Delta t)= 3$,
corresponding to panels (b) of Fig.~\ref{fig:conf2d},
where gravity effects are already visible in the system.
The observed data collapse onto a single master curve
suggests that the scaling behavior (\ref{eq:scaling}) holds reasonably well.
A slight deviation from the master curve can be detected at small
values of $k_y$ for the larger values of gravity. This is due to the presence
of liquid and vapor layers at walls extending over long space scales
so that the growth is no more self-similar.
Finally, the tail of the scaled structure factors follows
a power-law decay with an exponent equal to $-2$. This is consistent
with the Porod's law \cite{oono1988} $G(k_y,t) \sim k_y^{-(d+1)}$ 
where the dimension of the ordering space is $d=1$ in our case after
having averaged along the $k_x$-direction.
This latter result indicates that sharp interfaces are maintained along
the vertical direction despite the acceleration due to gravity.
A similar conclusion has been recently drawn in MD simulations of a
liquid-vapor system separating under gravity \cite{Davis2025}.

\begin{figure}[ht]
\begin{center}
\includegraphics*[width=0.9\columnwidth,angle=0]{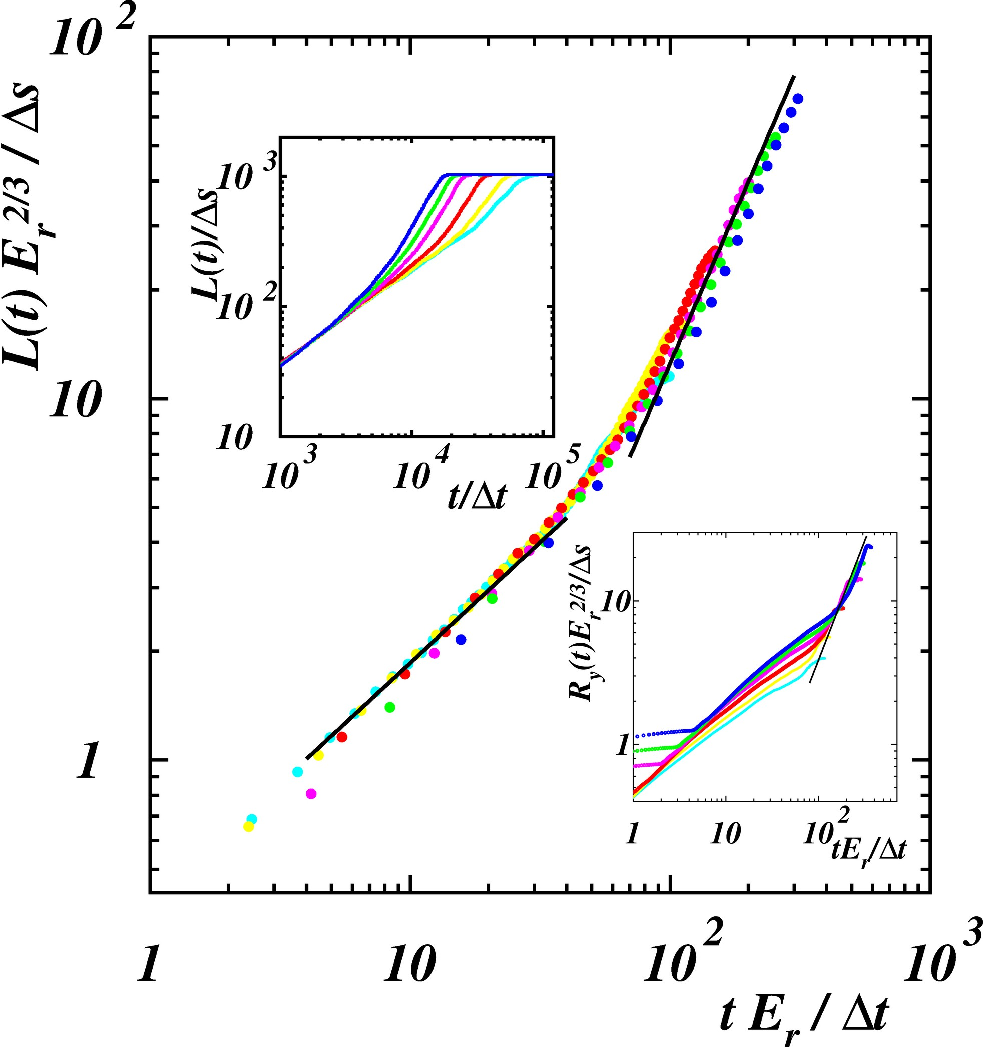}
\caption{Scaled thickness of layers at walls  $L(t) E_r^{2/3}/\Delta s$
  as a function of the scaled time $t E_r /\Delta t$
  in the cases with
  $E_r= 0.12 \times 10^{-2}$ (cyan symbols),  $0.20 \times 10^{-2}$ (yellow
  symbols),
  $0.41 \times 10^{-2}$ (red symbols), $0.82 \times 10^{-2}$ (purple symbols),
  $1.23 \times 10^{-2}$ (green symbols),
  $1.84 \times 10^{-2}$ (blue symbols). The black straight lines have slopes
  $0.66$ and $1.66$. Upper inset:  Thickness of layers at walls
  $L(t)/\Delta s$
  as a function of time $t/\Delta t$
  for the same cases of the main panel.
  Lower inset: Scaled size of domains along the vertical
  direction $R_y(t) E_r^{2/3}/\Delta s$
  as a function of the scaled time $t E_r /\Delta t$
  in the same cases of the main panel.
  The black straight line has slope $1.66$.
\label{fig:raf2}
}
\end{center}
\end{figure}
We consider the process of material accumulation
at walls. It is convenient to measure the average thickness
$L$ of the two layers forming
next to the walls \cite{Lacasta1993,articlecristeagravity}
\begin{equation}
  L(t)=\sum_{1 \leq x \leq N} \frac{\Big[y_b(x) + \big(H-y_t(x)\big)\Big]
    }{2N} \Delta s .
\end{equation} 
$y_b(x)$ and $y_t(x)$ are the $y$-coordinates
of all the lattice points,
where there is an interface between the liquid and vapor
phases, with the
minimum distances from the bottom and top walls, respectively, such that
$n_r(x,y_b(x))=n_r^{av}$ and $n_r(x,y_t(x))=n_r^{av}$.
The time evolution of $L(t)$ is shown in the upper inset
of Fig.~\ref{fig:raf2}. 
When phase separation begins, the layers at walls grow with no dependence
on the applied force following the time law $L(t) \sim t^{2/3}$. The
acceleration of the 
sedimentation of liquid at bottom promoted by gravity, determines
a faster growth rate being $L(t) \sim g t^{5/3}$.
At a later crossover time, which reduces with the gravity strength,
$L$ saturates at the maximum value $H/2$.
We find that numerical
data of $L$ before saturating, multiplied by $E_r^{2/3}$,
collapse onto a single curve when plotted as a function of
$t E_r$ over all the range of $E_r$ values (Fig.~\ref{fig:raf2}).
The master curve presents an initial slope $2/3$ followed by the faster
growth characterized by the exponent $5/3$.
The same scaling applies to $R_y(t)$ in the gravity-driven regime
as illustrated in the lower inset of Fig.~\ref{fig:raf2}.
The crossover to the the faster growth rate can be understood
  by introducing the capillary length $\Lambda_c$.
  This quantity is set by the surface tension and the gravity force.
  In the case of an interface separating two fluids,
  it results to be $\Lambda_c= \sqrt{\sigma/[(n_r^L - n_r^V) g]}$.
  Here $\sigma$ is the surface tension
  that in our model can be approximated as
  $\sigma= 2 \kappa (n_r^L - n_r^V)^2/(3 \chi)$ \cite{Negro2024}.
  A crossover to gravity-dominated motion occurs when $L(t) \gtrsim \Lambda_c$
  \cite{siggia1979}.
By considering in our study, for example, the cases at
$E_r= 0.12 \times 10^{-2}, 1.84 \times 10^{-2}$,
it is found that $\Lambda_c/\Delta s \sim 200, 50$,
respectively. These values are in
a very good agreement with the numerical values of $L(t)$
(see the cyan and blue lines in the upper inset of Fig.~\ref{fig:raf2})
where the crossover from the initial regime with exponent $2/3$
to the faster one is observed. Moreover, the late-time
growth exponent $5/3$
fits very well with the value $1.7 \pm 0.4$ found in experiments
of phase separation in lutidine water \cite{kim1978} which has been
attributed to gravity effects \cite{siggia1979}.

\begin{figure}[ht]
\begin{center}
\includegraphics*[width=0.9\columnwidth,angle=0]{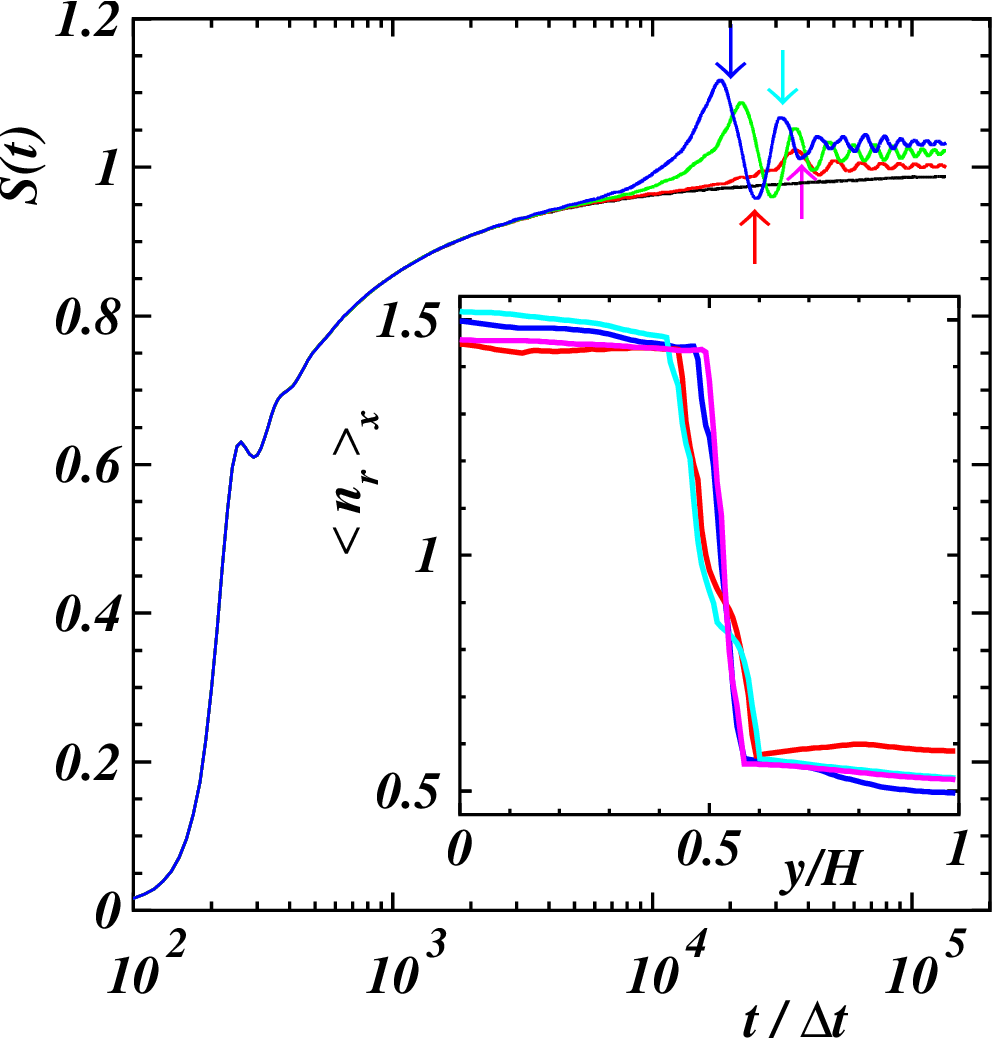}
\caption{Separation depth $S(t)$ in the cases with
  $E_r= 0$ (black line),  $0.41 \times
  10^{-2}$ (red line), $1.23 \times
  10^{-2}$ (green line), $1.84 \times
  10^{-2}$ (blue line). Inset: Profile of the density $n_r$, averaged along
  the horizontal direction, as a function of the vertical coordinate $y/H$
  in the case with $E_r=1.84 \times 10^{-2}$ at
  times $t/(5.2 \times 10^3 \Delta t)= 4$
  (blue line), $5$ (red line),
  $6$ (cyan line), $7$ (purple line), denoted by the corresponding colored
  arrows in the main panel. 
\label{fig:separation}
}
\end{center}
\end{figure}
Another possibility
to follow to evolution of the phase separation process is to keep
track of the distance of the local density from the
corresponding value at equilibrium. This can be done by introducing 
the separation depth $S(t)$ which is defined as
\cite{lamorgese2009}
\begin{equation}
  S(t)= \lla \frac{n_r({\mathbf r},t)-n_r^{av}}
  {n_r^{eq}({\mathbf r},t)-n_r^{av}} \rra
\label{eq:sep}  
\end{equation}
where the angular brackets denote an average over the whole system
and it is either $n_r^{eq}({\mathbf r})=n_r^L$ if
$n_r({\mathbf r})>n_r^{av}$ or
$n_r^{eq}({\mathbf r})=n_r^V$ if
$n_r({\mathbf r})<n_r^{av}$.
The time evolution of $S(t)$ is illustrated in Fig.~\ref{fig:separation}
for different values of $E_r$.
At the beginning, the phase separation occurs quickly and $S$ rapidly
increases up to $0.6$ suggesting that equilibrium is reached locally
once sharp interfaces separate different domains.
Later, the approach to the equilibrium proceeds
differently, depending on the value of gravity.
In the absence of gravity, the equilibrium at $S=1$ is reached smoothly
though with a slower rate than before. This is because single-phase
domains are characterized by small density gradients and densities
across interfaces change moderately.
By switching on the external force, two new features appear in the late time
behavior of $S(t)$.
The separation depth, after reaching an approximate value of $0.95$,
shows oscillations whose amplitude reduces in time while approaching
the asymptotic limit.
The oscillatory phenomenon
is similar to
the one previously observed in the
time behavior of $R_y(t)$ and manifests itself when the system is made of
two well separated domains of liquid and vapor at bottom and top, respectively,
of the system.
The amplitude of oscillations increases by strengthening the external force.
A last remark is related to the
final value of $S$ which increases with gravity becoming
larger than $1$.
This suggests, due to the definition (\ref{eq:sep}), that the system
can reach density values larger than $n_r^L$ and smaller than $n_r^V$ in
liquid and vapor domains, respectively, because of gravity. This aspect will
be later scrutinized. In order to understand the oscillations, we
consider the density profiles $\lla n_r({\mathbf r}) \rra_x$
along the vertical direction,
averaged horizontally. They are shown in the inset of 
Fig.~\ref{fig:separation} in the case of the largest value of gravity
at four consecutive times, approximately corresponding to the first two
maxima and two minima of the blue curve in the main panel.
It can be appreciated that after the formation of the two well separated
bands in the system, the approach to the final steady state occurs by
variations
of the densities in the bulk,  which oscillate around the
equilibrium values $n_r^L$ and $n_r^V$, while the position as well as the the
slope of the separating interfaces slightly change.

\subsection{Final state}

\begin{figure}[ht]
\begin{center}
\includegraphics*[width=0.9\columnwidth,angle=0]{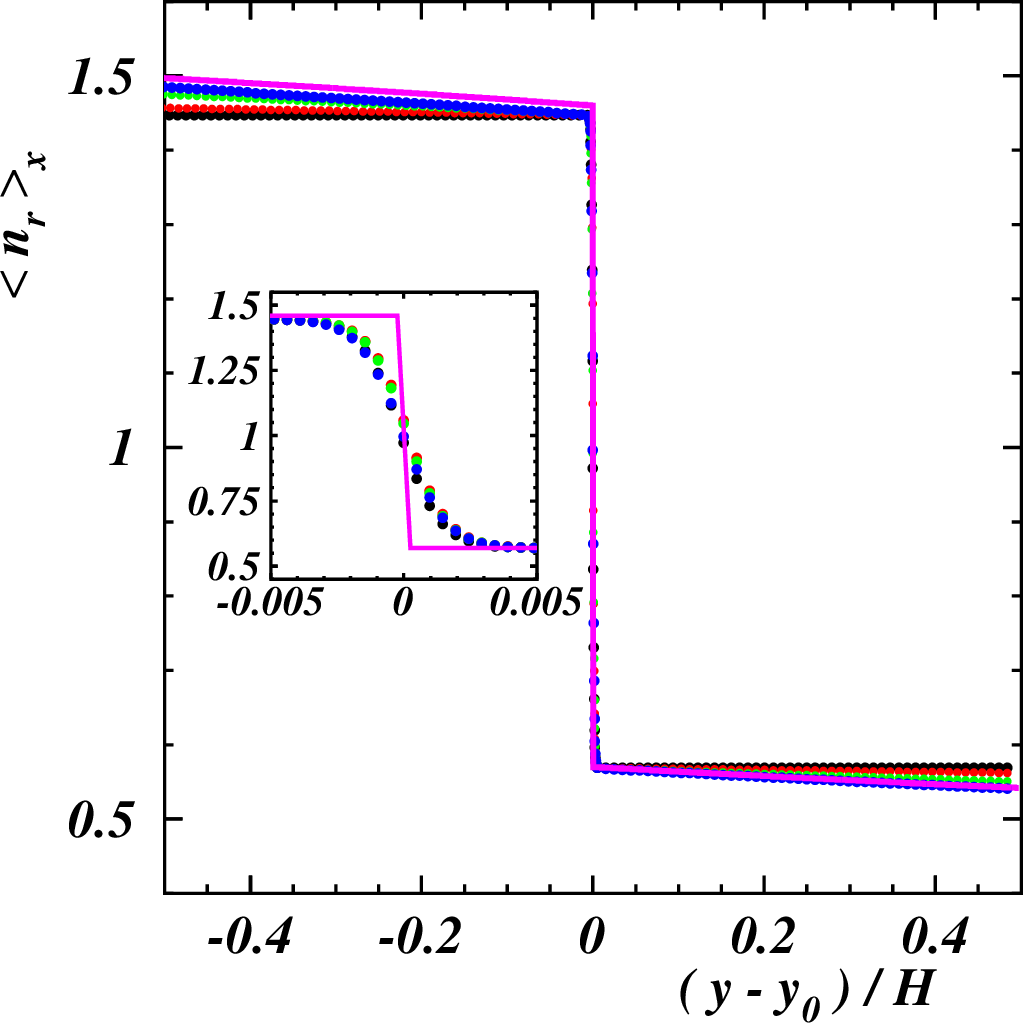}
\caption{Steady profiles of the density $n_r$, averaged along
  the horizontal direction, as a function of the vertical coordinate
  in the cases with $E_r= 0$ (black symbols),  $0.41 \times
  10^{-2}$ (red symbols), $1.23 \times
  10^{-2}$ (green symbols), $1.84 \times
  10^{-2}$ (blue symbols). The full purple line corresponds to the theoretical
  density profile, obtained from Eq.~(\ref{eq:ns1d}),
  with $E_r= 1.84 \times 10^{-2}$. Inset: Zoom of the density
  profile at the interface position $y-y_0=0$.
\label{fig:interfaccia}
}
\end{center}
\end{figure}
Once the phase separation has completed, the liquid sits completely
at the bottom and a single flat interface separates the two domains so that
the system is effectively one-dimensional.
Under these circumstances
there is no time dependence, the fluid velocity is zero so that 
it is possible to derive the density profile across the interface
\cite{Santos2002}.
The continuity equation (\ref{eq:cont}) is verified
and the Navier-Stokes equation (\ref{eq:ns})
becomes 
\begin{equation}
\partial_{y}\Pi_{yy}={\cal F}_{e,y}
\end{equation}
which can be written as
\begin{equation}
  \frac{d n_r}{d y}=
  -\frac{M g n_r / (R T_c)}{T_r/(1-n_r/3)^2-9 n_r/4} 
\label{eq:ns1d}
\end{equation}
with the use of Eq.~(\ref{eq:p}) and discarding the terms depending on
$\kappa$.
It is assumed that the interface separating the two phases is located at
the position $y=y_0$ so that the liquid and the vapor are
found between $0 \leq y \leq y_0$ and $y_0 < y \leq H$, respectively.
The values of densities at the boundary between the two phases are
the coexisting densities $n_r^L$ and $n_r^V$, as found with the Maxwell's rule,
at the temperature $T_r$ such that $p_r^{vdW}(n_r^L,T_r)=p_r^{vdW}(n_r^V,T_r)$.
By using these values of densities as initial conditions, Eq.~(\ref{eq:ns1d})
can be numerically integrated, starting from $y=y_0$, backward and forward
to derive the values of densities $n_r^l(y)$ and $n_r^v(y)$
in the liquid and vapor phases, respectively.
After this integration, the position $y_0$ of the interface can be obtained
numerically by
inserting the density profiles in the mass conservation equation
\cite{Santos2002}
\begin{equation}
\int_0^{y_0} dy n_r^l(y) + \int_{y_0}^H dy n_r^v(y)=n_r^{av} H .
\end{equation}

Numerical data of the interface profiles $\lla n_r \rra_x$
are plotted along the vertical
direction in
Fig.~\ref{fig:interfaccia} for different values of $E_r$.
The position $y_0$ is such that $\lla n_r \rra_x(y_0)=n_r^{av}$.
In the absence of gravity, densities are
constant within the two domains and equal
to the values $n_r^L$ and $n_r^V$. The effect of gravity is to
produce negative density gradients whose slopes decrease with gravity.
The net effect is that the values of liquid and vapor densities
in the bulk are
$n_{r, b}^l(y) > n_r^L$ for $0 \leq y \leq y_0$
and $n_{r, b}^v(y) < n_r^V$ for $y_0 < y \leq H$. This latter
result explains the fact that the separation depth $S$ is found to be larger
than $1$ in the long time limit. For the highest value of gravity we show
in the figure the theoretical density profile reconstructed according to the
aforementioned procedure. It can be observed that the agreement with simulation
results is excellent. The difference between the theoretical and numerical
values is less than $2\%$. A similar
deviation is observed in the numerical
pressure $\Pi_{yy}$ across the interface with respect to the theoretical values
so that thermodynamic consistency is still preserved within numerical accuracy.
It has to be remarked that the theoretical profile
is sharp while the one obtained by LB simulations is diffuse
by construction, as shown in the inset of Fig.~\ref{fig:interfaccia}.
Spurious velocities at the interface are negligible in the cases
both with and without gravity.

\begin{figure}[ht]
\begin{center}
\includegraphics*[width=0.9\columnwidth,angle=0]{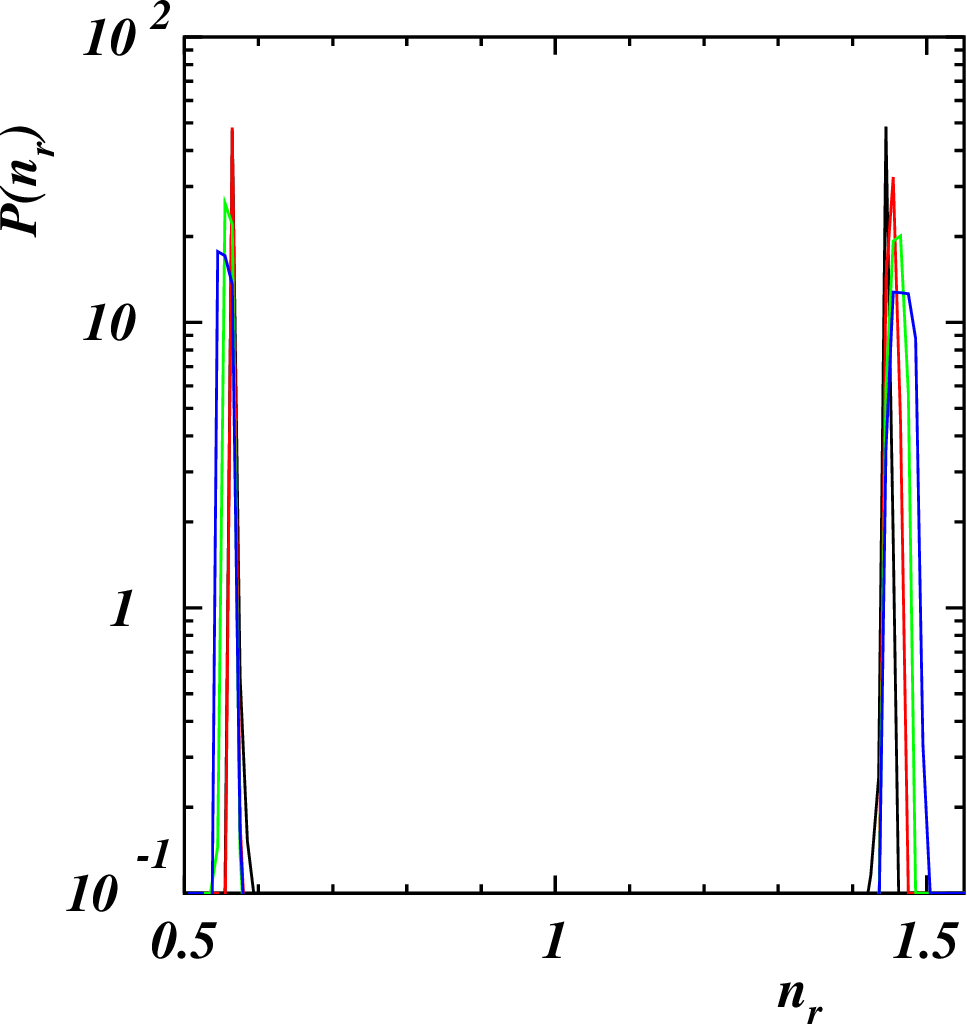}
\caption{Normalized distribution functions $P(n_r)$
  of the steady-state density $n_r$ in the cases with
  $E_r= 0$ (black line),  $0.41 \times
  10^{-2}$ (red line), $1.23 \times
  10^{-2}$ (green line), $1.84 \times
  10^{-2}$ (blue line).
\label{fig:pdf}
}
\end{center}
\end{figure}
In order to further characterize the density values when the sedimentation
process has completed, the 
normalized distribution functions $P(n_r)$ of density $n_r$ in the steady state
have been computed and depicted in Fig.~\ref{fig:pdf} for different values
of $E_r$. The distributions are bimodal: The peak of the liquid phase is
slightly higher
and narrower 
than the one of the vapor phase at fixed value of $E_r$.
Moreover, it can be seen that the two peaks broaden and decrease in height
when increasing gravity.

\section{Conclusions}
\label{sec:conclusions}

The phase separation process of a symmetric
liquid-vapor system confined between two horizontal walls,
suddenly quenched below the critical temperature,
has been investigated in a vertical gravity field
by extensive numerical simulations.
The equilibrium properties of the system are described by the van der Waals
equation of state and its dynamics has been considered by implementing
a lattice Boltzmann model. This approach allows us to correctly recover
the continuity and Navier-Stokes equations in order to properly
take into account hydrodynamic effects. These latter ones determine, in the
absence of gravity, a self-similar
coarsening dynamics with an exponent $2/3$,
  characteristic of an inertial regime.

The action of gravity produces a faster phase separation
leading to a complete segregation of liquid and vapor, as also observed
in MD simulations of nonideal fluids \cite{Davis2025}
and for binary fluid mixtures \cite{Badalassi2004,Bertei2022}.
By computing the structure factor, it has been possible to measure
the characteristic sizes of domains. We find that at late times
the growth
is accelerated along the gravitational direction with an effective
growth exponent which increases with gravity strength.
Along the transversal direction, the typical size increases very rapidly
after gravity induces accumulation of liquid and vapor at bottom and top
walls, respectively. The analysis of structure factor also enables to verify
that the Porod's law is still verified along the gravitational direction.
The average
thickness $L$ of accumulated material
at walls has also been measured. The initial behavior 
$L(t) \simeq t^{2/3}$ is followed
by a faster one at late times given by
$L(t) \simeq E_r t^{5/3}$, $E_r$ being the gravitational energy.
The final steady state is formed by two layers of liquid and vapor, completely
separated. The density profile perpendicular to the interface separating such
layers, depends on the gravity force and is found to be
in very good agreement with
the theoretical prediction.
Despite the present study has been performed in two dimensions, we
do not expect to observe different
growth exponents in three dimensions. In this latter case
it has been provided
convincing evidence \cite{Negro2024}
of the lack of the viscous regime with growth exponent $1$, attributed to the
Rayleigh-Plateau instability, differently from what observed
in the case of binary fluid mixtures.

The present study can provide a useful numerical tool to investigate
how different flow and force fields affect the dynamic behavior of
liquid-vapor systems
\cite{Zhang2008,wang2009,Lattuada2016,Gallo2024,Saiseau2024,Duraz2025,Uematsu2025}.
In particular, it might be worth to consider the
interplay of hydrodynamics and time-dependent force fields on coarsening
 \cite{pooley2004}.

\begin{acknowledgments}
This work was performed under the
auspices of GNFM-INdAM. The support of the National
Centre for HPC, Big Data, and Quantum Computing (Spoke 6, CN00000013)
is acknowledged.
\end{acknowledgments}

\section*{Data Availability Statement}

The data that support the ﬁndings of
this study are available from the
author upon reasonable
request.



\end{document}